\documentclass{article}
\usepackage{amsmath}
\begin{document}

\title{Calculation of critical index $\eta$ of the $\varphi^3$-theory in 4-loop approximation by the conformal bootstrap technique.}
\author{A. L. Pismensky \\
V. A. Fock Department of Theoretical Physics, Saint-Petersburg State University, \\
1, ul. Ulyanovskaya, 198504, Saint-Petersburg, Russia.}

\maketitle

\begin{abstract}
The method of calculation of $\varepsilon$-expansion in model of scalar field with $\varphi^3$-interaction based on conformal bootstrap equations is proposed. This technique is based on self-consistent skeleton equations involving full propagator and full triple vertex. Analytical computations of the Fisher's index $\eta$ are performed in four-loop approximation. The three-loop result coincides with one obtained previously by the renormalization group equations technique based on calculation of a larger number of Feynman diagrams. The four-loop result agrees with its numerical value obtained by other authors.
\end{abstract}

\section{Introduction.}
  Calculation of critical exponents is one of the main problems in the theory of critical behavior. One of the very effective ways of its solving is the renormalization group equations technique. It enables us to construct $\varepsilon$-expansions, that is, power series in deviation $\varepsilon$ from the logarithmic dimension. Another method which is used for description of the system behavior at critical points is the self-consistent equations which result from discarding all the bare contributions in the skeleton equations for Green's functions [1---5]. Is was used for construction of $1/n$-expansions in the $O(n)$-symmetric $(\varphi^2)^2$ model. The main advantage of this method is the significant reduction in the number of Feynman's diagrams required for solving of the problem. The self-consistent equations for full propagator and full triple vertex are the base of the conformal bootstrap technique which was used for calculation of $1/n^3$-correction to the index $\eta$ in $(\varphi^2)^2$ model \cite{5}. In the present paper, we propose to use this method for construction of the $\varepsilon$-expansion in $\varphi^3$ model and compute the $\varepsilon^4$-correction to the index $\eta$. For calculation of the Feynman's graphs with necessary accuracy we use the methods described in [1, 6---9].

\section{Conformal bootstrap method for $\varphi^3$-theory.}
We investigate the use of conformal bootstrap technique for the simplest massless $\varphi^3$ theory with one scalar field $\varphi(x)$:
$$ {\cal L} = \frac{1}{2} (\partial \varphi)^2 + \frac{\lambda}{3!} \varphi^3 $$
in the Euclidian space of the dimension $d=6+2\varepsilon$.
The propagator and the vertex function in non-logarithmic dimension ($\varepsilon \neq 0$) are known to be the power functions of coordinates:
$$ D(x_1,x_2) = \frac{A}{(x_1-x_2)^{2\alpha}}, \ \Gamma(x_1, x_2, x_3) = \frac{C}{(x_1-x_2)^{2a}(x_1-x_3)^{2a}(x_2-x_3)^{2a}} $$
where $ \alpha = \frac{d}{2} - 1 + \frac{\eta}{2} $, \ $\eta$ is a critical exponent, $a=\frac{d-\alpha}{2}$.
Our aim is to find $\eta$ in the form of $\varepsilon$-expansion up to $\varepsilon^4$:
$$\eta = \eta_1 \varepsilon + \eta_2 \varepsilon^2 + \eta_3 \varepsilon^3 + \eta_4 \varepsilon^4 + {\cal O}(\varepsilon^5).$$

The system of equations of the conformal bootstrap is the following \cite{1,5}:
\begin{equation}
\label{bootsys}
\left\{ \begin{array}{l}
\left. V(\alpha, u; \omega) \right|_{\omega=0} = 1, \\
\left. 2 p(\alpha) = u S(\alpha) \frac{\partial V(\alpha, u; \omega)}{\partial \omega} \right|_{\omega=0}
\end{array} \right.
\end{equation}
where $ p(\alpha) = \pi^{-d} H(\alpha{-}d/2, d/2{-}\alpha),\ S(\alpha)=\pi^{2d} \frac{H(\alpha, \alpha, \alpha, a, a, a, d/2{+}a{-}\alpha)}{\Gamma(d/2)}, $
$H(z) = \frac{\Gamma(z')}{\Gamma(z)}$,\ $\Gamma(z)$ is the Euler's gamma function, $z'=d/2-z $, \
$ H(z_1, z_2, z_3, ...) = H(z_1) H(z_2) H(z_3) ... $,
$u=C^2A^3$ (a renorm-invariant combination of amplitudes). \\
The function $V(\alpha, u; \omega)$ is defined by the condition:
\begin{center}
\begin{picture}(215,40)
\put(0,22){\scriptsize $\omega$}
\put(0,20){\line(1,0){5}}
\put(8,20){\circle{6}}
\put(10,22){\line(1,1){11}}
\put(10,18){\line(1,-1){11}}
\put(23,35){\circle{6}}
\put(23,5){\circle{6}}
\put(23,8){\line(0,1){24}}
\put(26,5){\line(1,0){5}}
\put(26,35){\line(1,0){5}}
\put(32,18){$+\frac{1}{2}$}
\put(50,22){\scriptsize $\omega$}
\put(50,20){\line(1,0){5}}
\put(58,20){\circle{6}}
\put(60,22){\line(1,1){11}}
\put(60,18){\line(1,-1){11}}
\put(73,35){\circle{6}}
\put(73,5){\circle{6}}
\put(76,5){\line(1,0){24}}
\put(76,35){\line(1,0){24}}
\put(103,35){\circle{6}}
\put(103,5){\circle{6}}
\put(106,5){\line(1,0){5}}
\put(106,35){\line(1,0){5}}
\put(75,7){\line(1,1){26}}
\put(75,33){\line(1,-1){26}}
\put(112,18){$+... =V(\alpha,u;\omega)$}
\put(192,22){\scriptsize $\omega$}
\put(192,20){\line(1,0){5}}
\put(200,20){\circle{6}}
\qbezier(202,22)(204,24)(206,26)
\qbezier(202,18)(204,16)(206,14)
\end{picture}
\end{center}
where circle is vertex function, the circle marked with $\omega$ is the regularized vertex function: \\
\begin{center}
\begin{picture}(180,30)
\put(0,15){\line(1,0){5}}
\put(8,15){\circle{6}}
\qbezier(10,17)(12,19)(14,21)
\qbezier(10,13)(12,11)(14,9)
\put(20,13){$=$}
\put(30,17){\scriptsize $\alpha$}
\put(30,15){\line(1,0){5}}
\put(35,15){\circle*{2}}
\put(35,15){\line(2,1){20}}
\put(35,15){\line(2,-1){20}}
\put(55,5){\circle*{2}}
\put(55,25){\circle*{2}}
\put(55,5){\line(0,1){20}}
\put(55,5){\line(1,0){5}}
\put(55,25){\line(1,0){5}}
\put(55,7){\scriptsize $\alpha$}
\put(55,27){\scriptsize $\alpha$}
\put(55,15){\scriptsize $a$}
\put(43,22){\scriptsize $a$}
\put(43,12){\scriptsize $a$}
\put(65,10){,}
\put(80,17){\scriptsize $\omega$}
\put(80,15){\line(1,0){5}}
\put(88,15){\circle{6}}
\qbezier(90,17)(92,19)(94,21)
\qbezier(90,13)(92,11)(94,9)
\put(100,13){$=$}
\put(110,17){\scriptsize $\alpha+2\omega$}
\put(130,15){\line(1,0){5}}
\put(135,15){\circle*{2}}
\put(135,15){\line(2,1){20}}
\put(135,15){\line(2,-1){20}}
\put(155,5){\circle*{2}}
\put(155,25){\circle*{2}}
\put(155,5){\line(0,1){20}}
\put(155,5){\line(1,0){5}}
\put(155,25){\line(1,0){5}}
\put(155,7){\scriptsize $\alpha$}
\put(155,27){\scriptsize $\alpha$}
\put(155,15){\scriptsize $a+\omega$}
\put(133,22){\scriptsize $a-\omega$}
\put(133,2){\scriptsize $a-\omega$}
\end{picture}
\end{center}

For calculation of $\eta_4$ we need to take into consideration the following diagrams:
\begin{center}
\begin{picture}(500,385)
\put(0,348){$V(\alpha,u;\omega)$}
\put(50,352){\scriptsize $\omega$}
\put(50,350){\line(1,0){5}}
\put(58,350){\circle{6}}
\qbezier(60,352)(62,354)(64,356)
\qbezier(60,348)(62,346)(64,344)
\put(67,348){$=$}

\put(80,352){\scriptsize $\omega$}
\put(80,350){\line(1,0){5}}
\put(88,350){\circle{6}}
\put(90,352){\line(1,1){11}}
\put(90,348){\line(1,-1){11}}
\put(103,365){\circle{6}}
\put(103,335){\circle{6}}
\put(103,338){\line(0,1){24}}
\put(106,335){\line(1,0){5}}
\put(106,365){\line(1,0){5}}
\put(95,323){$\gamma_1$}

\put(112,348){$+\frac{1}{2}$}
\put(130,352){\scriptsize $\omega$}
\put(130,350){\line(1,0){5}}
\put(138,350){\circle{6}}
\put(140,352){\line(1,1){11}}
\put(140,348){\line(1,-1){11}}
\put(153,365){\circle{6}}
\put(153,335){\circle{6}}
\put(156,335){\line(1,0){24}}
\put(156,365){\line(1,0){24}}
\put(183,365){\circle{6}}
\put(183,335){\circle{6}}
\put(186,335){\line(1,0){5}}
\put(186,365){\line(1,0){5}}
\put(155,337){\line(1,1){26}}
\put(155,363){\line(1,-1){26}}
\put(165,323){$\gamma_2$}

\put(192,348){$+$}
\put(205,352){\scriptsize $\omega$}
\put(205,350){\line(1,0){5}}
\put(213,350){\circle{6}}
\put(215,352){\line(1,1){26}}
\put(215,348){\line(1,-1){26}}
\put(243,380){\circle{6}}
\put(243,320){\circle{6}}
\put(243,350){\circle{6}}
\put(273,320){\circle{6}}
\put(273,350){\circle{6}}
\put(273,380){\circle{6}}
\put(246,320){\line(1,0){24}}
\put(246,350){\line(1,0){24}}
\put(246,380){\line(1,0){24}}
\put(243,323){\line(0,1){24}}
\put(243,353){\line(0,1){24}}
\put(273,323){\line(0,1){24}}
\put(273,353){\line(0,1){24}}
\put(276,320){\line(1,0){5}}
\put(276,380){\line(1,0){5}}
\put(250,308){$\gamma_{31}$}

\put(285,348){$+$}
\put(300,352){\scriptsize $\omega$}
\put(300,350){\line(1,0){5}}
\put(308,350){\circle{6}}
\put(310,352){\line(1,1){26}}
\put(310,348){\line(1,-1){26}}
\put(338,380){\circle{6}}
\put(338,320){\circle{6}}
\put(338,350){\circle{6}}
\put(368,320){\circle{6}}
\put(368,350){\circle{6}}
\put(368,380){\circle{6}}
\put(341,320){\line(1,0){24}}
\put(340,352){\line(1,1){26}}
\put(340,378){\line(1,-1){26}}
\put(338,323){\line(0,1){24}}
\put(338,353){\line(0,1){24}}
\put(368,323){\line(0,1){24}}
\put(368,353){\line(0,1){24}}
\put(371,320){\line(1,0){5}}
\put(371,380){\line(1,0){5}}
\put(345,308){$\gamma_{32a}$}

\put(385,348){$+$}
\put(400,352){\scriptsize $\omega$}
\put(400,350){\line(1,0){5}}
\put(408,350){\circle{6}}
\put(410,352){\line(1,1){26}}
\put(410,348){\line(1,-1){26}}
\put(438,380){\circle{6}}
\put(438,320){\circle{6}}
\put(438,350){\circle{6}}
\put(468,320){\circle{6}}
\put(468,350){\circle{6}}
\put(468,380){\circle{6}}
\put(441,380){\line(1,0){24}}
\put(440,322){\line(1,1){26}}
\put(440,348){\line(1,-1){26}}
\put(438,323){\line(0,1){24}}
\put(438,353){\line(0,1){24}}
\put(468,323){\line(0,1){24}}
\put(468,353){\line(0,1){24}}
\put(471,320){\line(1,0){5}}
\put(471,380){\line(1,0){5}}
\put(445,308){$\gamma_{32b}$}

\put(486,348){$+$}

\put(0,248){$+$}
\put(15,252){\scriptsize $\omega$}
\put(15,250){\line(1,0){5}}
\put(23,250){\circle{6}}
\put(25,252){\line(1,1){26}}
\put(25,248){\line(1,-1){26}}
\put(53,280){\circle{6}}
\put(53,220){\circle{6}}
\put(53,250){\circle{6}}
\put(83,220){\circle{6}}
\put(83,250){\circle{6}}
\put(83,280){\circle{6}}
\put(56,220){\line(1,0){24}}
\put(56,250){\line(1,0){24}}
\put(56,280){\line(1,0){24}}
\put(55,252){\line(1,1){26}}
\put(55,278){\line(1,-1){26}}
\put(53,223){\line(0,1){24}}
\put(83,223){\line(0,1){24}}
\put(86,220){\line(1,0){5}}
\put(86,280){\line(1,0){5}}
\put(60,208){$\gamma_{32c}$}

\put(100,248){$+3$}
\put(120,252){\scriptsize $\omega$}
\put(120,250){\line(1,0){5}}
\put(128,250){\circle{6}}
\put(130,252){\line(1,1){11}}
\put(130,248){\line(1,-1){11}}
\put(143,265){\circle{6}}
\put(143,235){\circle{6}}
\put(145,267){\line(1,1){11}}
\put(145,233){\line(1,-1){11}}
\put(158,280){\circle{6}}
\put(158,220){\circle{6}}
\put(188,280){\circle{6}}
\put(188,220){\circle{6}}
\put(203,265){\circle{6}}
\put(203,235){\circle{6}}
\put(161,220){\line(1,0){24}}
\put(161,280){\line(1,0){24}}
\put(190,278){\line(1,-1){11}}
\put(190,222){\line(1,1){11}}
\put(159,222){\line(1,2){28}}
\put(159,278){\line(1,-2){28}}
\put(145,264){\line(2,-1){56}}
\put(145,236){\line(2,1){56}}
\put(206,235){\line(1,0){5}}
\put(206,265){\line(1,0){5}}
\put(165,208){$\gamma_{41}$}

\put(220,248){$+3$}
\put(240,252){\scriptsize $\omega$}
\put(240,250){\line(1,0){5}}
\put(248,250){\circle{6}}
\put(250,252){\line(1,1){11}}
\put(250,248){\line(1,-1){11}}
\put(263,265){\circle{6}}
\put(263,235){\circle{6}}
\put(265,267){\line(1,1){11}}
\put(265,233){\line(1,-1){11}}
\put(278,280){\circle{6}}
\put(278,220){\circle{6}}
\put(308,280){\circle{6}}
\put(308,220){\circle{6}}
\put(323,265){\circle{6}}
\put(323,235){\circle{6}}
\put(281,220){\line(1,0){24}}
\put(281,280){\line(1,0){24}}
\put(310,278){\line(1,-1){11}}
\put(310,222){\line(1,1){11}}
\put(278,223){\line(0,1){54}}
\put(308,223){\line(0,1){54}}
\put(265,264){\line(2,-1){56}}
\put(265,236){\line(2,1){56}}
\put(326,235){\line(1,0){5}}
\put(326,265){\line(1,0){5}}
\put(285,208){$\gamma_{42}$}

\put(340,248){$+6$}
\put(360,252){\scriptsize $\omega$}
\put(360,250){\line(1,0){5}}
\put(368,250){\circle{6}}
\put(370,252){\line(1,1){11}}
\put(370,248){\line(1,-1){11}}
\put(383,265){\circle{6}}
\put(383,235){\circle{6}}
\put(385,267){\line(1,1){11}}
\put(385,233){\line(1,-1){11}}
\put(398,280){\circle{6}}
\put(398,220){\circle{6}}
\put(428,280){\circle{6}}
\put(428,220){\circle{6}}
\put(443,265){\circle{6}}
\put(443,235){\circle{6}}
\put(401,220){\line(1,0){24}}
\put(401,280){\line(1,0){24}}
\put(430,278){\line(1,-1){11}}
\put(430,222){\line(1,1){11}}
\put(399,222){\line(1,2){28}}
\put(400,278){\line(1,-1){41}}
\put(385,263){\line(1,-1){41}}
\put(385,236){\line(2,1){56}}
\put(446,235){\line(1,0){5}}
\put(446,265){\line(1,0){5}}
\put(410,208){$\gamma_{43}$}

\put(460,248){$+$}

\put(0,148){$+6$}
\put(20,152){\scriptsize $\omega$}
\put(20,150){\line(1,0){5}}
\put(28,150){\circle{6}}
\put(30,152){\line(1,1){26}}
\put(30,148){\line(1,-1){26}}
\put(58,180){\circle{6}}
\put(58,120){\circle{6}}
\put(88,180){\circle{6}}
\put(118,180){\circle{6}}
\put(58,150){\circle{6}}
\put(88,150){\circle{6}}
\put(118,150){\circle{6}}
\put(118,120){\circle{6}}
\put(61,180){\line(1,0){24}}
\put(91,180){\line(1,0){24}}
\put(61,150){\line(1,0){24}}
\put(91,150){\line(1,0){24}}
\put(61,120){\line(1,0){54}}
\put(58,123){\line(0,1){24}}
\put(118,123){\line(0,1){24}}
\put(88,153){\line(0,1){24}}
\put(60,151){\line(2,1){56}}
\put(60,179){\line(2,-1){56}}
\put(121,120){\line(1,0){5}}
\put(121,180){\line(1,0){5}}
\put(80,108){$\gamma_{44}$}

\put(130,148){$+3$}
\put(150,152){\scriptsize $\omega$}
\put(150,150){\line(1,0){5}}
\put(158,150){\circle{6}}
\put(159,152){\line(1,2){13}}
\put(159,148){\line(1,-2){13}}
\put(173,180){\circle{6}}
\put(173,120){\circle{6}}
\put(188,165){\circle{6}}
\put(188,135){\circle{6}}
\put(218,165){\circle{6}}
\put(218,135){\circle{6}}
\put(233,180){\circle{6}}
\put(233,120){\circle{6}}
\put(175,122){\line(1,1){11}}
\put(175,178){\line(1,-1){11}}
\put(191,135){\line(1,0){24}}
\put(191,165){\line(1,0){24}}
\put(188,138){\line(0,1){24}}
\put(218,138){\line(0,1){24}}
\put(220,167){\line(1,1){11}}
\put(220,133){\line(1,-1){11}}
\put(176,180){\line(1,0){54}}
\put(176,120){\line(1,0){54}}
\put(236,120){\line(1,0){5}}
\put(236,180){\line(1,0){5}}
\put(195,108){$\gamma_{45}$}

\put(240,148){$+\frac{3}{2}$}
\put(260,152){\scriptsize $\omega$}
\put(260,150){\line(1,0){5}}
\put(268,150){\circle{6}}
\put(270,152){\line(1,1){11}}
\put(270,148){\line(1,-1){11}}
\put(283,165){\circle{6}}
\put(283,135){\circle{6}}
\put(285,167){\line(1,1){11}}
\put(285,133){\line(1,-1){11}}
\put(298,180){\circle{6}}
\put(298,120){\circle{6}}
\put(328,180){\circle{6}}
\put(328,120){\circle{6}}
\put(343,165){\circle{6}}
\put(343,135){\circle{6}}
\put(301,120){\line(1,0){24}}
\put(301,180){\line(1,0){24}}
\put(330,178){\line(1,-1){11}}
\put(330,122){\line(1,1){11}}
\put(285,137){\line(1,1){41}}
\put(300,122){\line(1,1){41}}
\put(285,163){\line(1,-1){41}}
\put(300,178){\line(1,-1){41}}
\put(346,135){\line(1,0){5}}
\put(346,165){\line(1,0){5}}
\put(305,108){$\gamma_{46}$}

\put(360,148){$+3$}
\put(380,152){\scriptsize $\omega$}
\put(380,150){\line(1,0){5}}
\put(388,150){\circle{6}}
\put(390,152){\line(1,1){26}}
\put(390,148){\line(1,-1){26}}
\put(418,180){\circle{6}}
\put(418,120){\circle{6}}
\put(448,180){\circle{6}}
\put(478,180){\circle{6}}
\put(418,150){\circle{6}}
\put(448,150){\circle{6}}
\put(478,150){\circle{6}}
\put(478,120){\circle{6}}
\put(421,180){\line(1,0){24}}
\put(451,180){\line(1,0){24}}
\put(421,150){\line(1,0){24}}
\put(451,150){\line(1,0){24}}
\put(421,120){\line(1,0){54}}
\put(418,123){\line(0,1){24}}
\put(478,123){\line(0,1){24}}
\put(420,178){\line(1,-1){26}}
\put(450,178){\line(1,-1){26}}
\put(420,151){\line(2,1){56}}
\put(481,120){\line(1,0){5}}
\put(481,180){\line(1,0){5}}
\put(440,108){$\gamma_{47}$}

\put(490,148){$+$}

\put(0,48){$+3$}
\put(20,52){\scriptsize $\omega$}
\put(20,50){\line(1,0){5}}
\put(28,50){\circle{6}}
\put(30,52){\line(1,1){26}}
\put(30,49){\line(1,-1){26}}
\put(58,80){\circle{6}}
\put(58,20){\circle{6}}
\put(88,80){\circle{6}}
\put(118,80){\circle{6}}
\put(58,50){\circle{6}}
\put(88,50){\circle{6}}
\put(118,50){\circle{6}}
\put(118,20){\circle{6}}
\put(61,80){\line(1,0){24}}
\put(91,80){\line(1,0){24}}
\put(61,50){\line(1,0){24}}
\put(91,50){\line(1,0){24}}
\put(61,20){\line(1,0){54}}
\put(58,23){\line(0,1){24}}
\put(118,23){\line(0,1){24}}
\put(60,52){\line(1,1){26}}
\put(90,52){\line(1,1){26}}
\put(60,79){\line(2,-1){56}}
\put(121,20){\line(1,0){5}}
\put(121,80){\line(1,0){5}}
\put(80,8){$\gamma_{48}$}

\put(130,48){$+ \frac{1}{2}$}
\put(150,52){\scriptsize $\omega$}
\put(150,50){\line(1,0){5}}
\put(158,50){\circle{6}}
\put(160,51){\line(2,1){26}}
\put(160,49){\line(2,-1){26}}
\put(188,65){\circle{6}}
\put(188,35){\circle{6}}
\put(218,80){\circle{6}}
\put(218,20){\circle{6}}
\put(248,95){\circle{6}}
\put(248,5){\circle{6}}
\put(248,35){\circle{6}}
\put(248,65){\circle{6}}
\put(190,66){\line(2,1){26}}
\put(220,81){\line(2,1){26}}
\put(190,34){\line(2,-1){26}}
\put(220,19){\line(2,-1){26}}
\put(190,36){\line(2,1){56}}
\put(190,64){\line(2,-1){56}}
\put(248,8){\line(0,1){24}}
\put(248,38){\line(0,1){24}}
\put(248,68){\line(0,1){24}}
\put(218,23){\line(0,1){54}}
\put(251,5){\line(1,0){5}}
\put(251,95){\line(1,0){5}}
\put(190,8){$\gamma_{49}$}
\put(260,48){$+\ ...$}
\end{picture}
\end{center}
We have
$$ V = \gamma_1 u + \frac{1}{2} \gamma_2 u^2 + (\gamma_{31} + \gamma_{32a} + \gamma_{32b} + \gamma_{32c}) u^3 + (3\gamma_{41}+3\gamma_{42}+6\gamma_{43}+6\gamma_{44}+3\gamma_{45}+\frac{3}{2}\gamma_{46}+3\gamma_{47}+3\gamma_{48}+\frac{1}{2}\gamma_{49}) u^4 +... $$
We chose the dimension $d=6+2\varepsilon$ because all these diagrams converge with $d>6$ only.

There are 3 major formulas of calculation of massless diagrams: \\
\begin{picture}(460,35)
\put(0,13){loop:}
\put(30,15){\circle*{2}}
\put(70,15){\circle*{2}}
\qbezier(30,15)(50,30)(70,15)
\qbezier(30,15)(50,0)(70,15)
\put(75,13){$=$}
\put(90,15){\circle*{2}}
\put(130,15){\circle*{2}}
\put(90,15){\line(1,0){40}}
\put(48,25){\scriptsize $z_1$}
\put(48,2){\scriptsize $z_2$}
\put(99,17){\scriptsize $z_1+z_2$}
\put(134,13){,}
\put(150,13){chain:}
\put(185,15){\circle*{2}}
\put(225,15){\circle*{2}}
\put(265,15){\circle*{2}}
\put(185,15){\line(1,0){80}}
\put(203,17){\scriptsize $z_1$}
\put(243,17){\scriptsize $z_2$}
\put(270,13){$=\pi^{d/2}H(z_1,z_2,d-z_1-z_2)$}
\put(395,15){\circle*{2}}
\put(450,15){\circle*{2}}
\put(395,15){\line(1,0){55}}
\put(398,17){\scriptsize $z_1+z_2-d/2$}
\put(454,13){,}
\end{picture} \\
\begin{picture}(380,40)
\put(0,18){unique triple vertex:}
\put(95,5){\circle*{2}}
\put(135,5){\circle*{2}}
\put(115,35){\circle*{2}}
\put(115,15){\circle*{2}}
\put(95,5){\line(2,1){20}}
\put(115,15){\line(2,-1){20}}
\put(115,15){\line(0,1){20}}
\put(103,13){\scriptsize $z_1$}
\put(123,13){\scriptsize $z_2$}
\put(116,25){\scriptsize $z_3$}
\put(135,18){$=\pi^{d/2}H(z_1,z_2,z_3)$}
\put(227,5){\circle*{2}}
\put(267,5){\circle*{2}}
\put(247,35){\circle*{2}}
\put(227,5){\line(1,0){40}}
\put(227,5){\line(2,3){20}}
\put(247,35){\line(2,-3){20}}
\put(260,18){\scriptsize $z_1'$}
\put(226,18){\scriptsize $z_2'$}
\put(244,7){\scriptsize $z_3'$}
\put(280,18){with\ \ $z_1 + z_2 + z_3 = d$}
\end{picture} \\
\begin{picture}(210,20)
\put(0,8){where line with index $z$ is:}
\put(125,10){\circle*{2}}
\put(155,10){\circle*{2}}
\put(125,10){\line(1,0){30}}
\put(138,12){\scriptsize $z$}
\put(120,3){$x_1$}
\put(150,3){$x_2$}
\put(160,8){$=\frac{1}{(x_1-x_2)^{2z}},$}
\end{picture} \\
\begin{picture}(350,20)
\put(0,8){if the index of line is not specified, that it is equal to unity:}
\put(270,10){\circle*{2}}
\put(300,10){\circle*{2}}
\put(270,10){\line(1,0){30}}
\put(265,3){$x_1$}
\put(295,3){$x_2$}
\put(305,8){$=\frac{1}{(x_1-x_2)^{2}}.$}
\end{picture}

Using the inversion transformation we reduce three-tail diagrams to two-tail ones. We illustrate this with the example of $\gamma_1$.
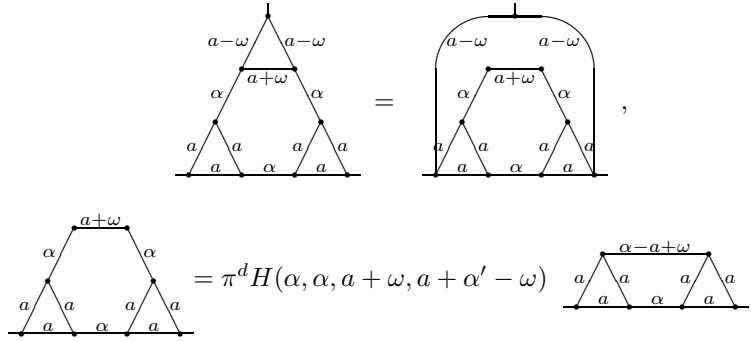
\begin{center}
\begin{picture}(90,70)
\put(0,5){\line(1,0){70}}
\put(5,5){\circle*{2}}
\put(25,5){\circle*{2}}
\put(45,5){\circle*{2}}
\put(65,5){\circle*{2}}
\put(5,5){\line(1,2){30}}
\put(35,65){\line(1,-2){30}}
\put(15,25){\circle*{2}}
\put(25,45){\circle*{2}}
\put(35,65){\circle*{2}}
\put(45,45){\circle*{2}}
\put(55,25){\circle*{2}}
\put(35,65){\line(0,1){5}}
\put(15,25){\line(1,-2){10}}
\put(45,5){\line(1,2){10}}
\put(25,45){\line(1,0){20}}
\put(33,6){\scriptsize $\alpha$}
\put(13,34){\scriptsize $\alpha$}
\put(51,34){\scriptsize $\alpha$}
\put(13,6){\scriptsize $a$}
\put(53,6){\scriptsize $a$}
\put(4,14){\scriptsize $a$}
\put(44,14){\scriptsize $a$}
\put(21,14){\scriptsize $a$}
\put(61,14){\scriptsize $a$}
\put(41,54){\scriptsize $a{-}\omega$}
\put(12,54){\scriptsize $a{-}\omega$}
\put(27,40){\scriptsize $a{+}\omega$}
\put(75,30){$=$}
\end{picture}
\begin{picture}(80,70)
\put(0,5){\line(1,0){70}}
\put(5,5){\circle*{2}}
\put(25,5){\circle*{2}}
\put(45,5){\circle*{2}}
\put(65,5){\circle*{2}}
\put(5,5){\line(1,2){20}}
\put(45,45){\line(1,-2){20}}
\put(15,25){\circle*{2}}
\put(25,45){\circle*{2}}
\put(35,65){\circle*{2}}
\put(45,45){\circle*{2}}
\put(55,25){\circle*{2}}
\put(35,65){\line(0,1){5}}
\put(15,25){\line(1,-2){10}}
\put(45,5){\line(1,2){10}}
\put(25,45){\line(1,0){20}}
\put(33,6){\scriptsize $\alpha$}
\put(13,34){\scriptsize $\alpha$}
\put(51,34){\scriptsize $\alpha$}
\put(13,6){\scriptsize $a$}
\put(53,6){\scriptsize $a$}
\put(4,14){\scriptsize $a$}
\put(44,14){\scriptsize $a$}
\put(21,14){\scriptsize $a$}
\put(61,14){\scriptsize $a$}
\put(44,54){\scriptsize $a{-}\omega$}
\put(9,54){\scriptsize $a{-}\omega$}
\put(27,40){\scriptsize $a{+}\omega$}
\put(35,5){\oval(60,120)[t]}
\put(75,30){$,$}
\end{picture}
\end{center}

\begin{center}
\begin{picture}(300,50)
\put(0,5){\line(1,0){70}}
\put(5,5){\circle*{2}}
\put(25,5){\circle*{2}}
\put(45,5){\circle*{2}}
\put(65,5){\circle*{2}}
\put(5,5){\line(1,2){20}}
\put(45,45){\line(1,-2){20}}
\put(15,25){\circle*{2}}
\put(25,45){\circle*{2}}
\put(45,45){\circle*{2}}
\put(55,25){\circle*{2}}
\put(15,25){\line(1,-2){10}}
\put(45,5){\line(1,2){10}}
\put(25,45){\line(1,0){20}}
\put(33,6){\scriptsize $\alpha$}
\put(13,34){\scriptsize $\alpha$}
\put(51,34){\scriptsize $\alpha$}
\put(13,6){\scriptsize $a$}
\put(53,6){\scriptsize $a$}
\put(4,14){\scriptsize $a$}
\put(44,14){\scriptsize $a$}
\put(21,14){\scriptsize $a$}
\put(61,14){\scriptsize $a$}
\put(27,46){\scriptsize $a{+}\omega$}
\put(70,23){$=\pi^d H(\alpha,\alpha,a+\omega,a+\alpha'-\omega)$}
\put(210,15){\line(1,0){70}}
\put(215,15){\circle*{2}}
\put(235,15){\circle*{2}}
\put(255,15){\circle*{2}}
\put(275,15){\circle*{2}}
\put(215,15){\line(1,2){10}}
\put(265,35){\line(1,-2){10}}
\put(225,35){\circle*{2}}
\put(265,35){\circle*{2}}
\put(225,35){\line(1,-2){10}}
\put(255,15){\line(1,2){10}}
\put(225,35){\line(1,0){40}}
\put(243,16){\scriptsize $\alpha$}
\put(223,16){\scriptsize $a$}
\put(263,16){\scriptsize $a$}
\put(214,24){\scriptsize $a$}
\put(254,24){\scriptsize $a$}
\put(231,24){\scriptsize $a$}
\put(271,24){\scriptsize $a$}
\put(231,36){\scriptsize $\alpha{-}a{+}\omega$}
\end{picture}
\end{center}
We denote:
$$ \Psi (\alpha; \omega) \equiv \pi^d H(\alpha,\alpha,a+\omega,a+\alpha'-\omega). $$
We have
$$ \gamma_1 = \Psi \gamma_1'; \ \frac{\partial \gamma_1}{\partial \omega} = \frac{\partial \Psi}{\partial \omega} \gamma_1' + \Psi \frac{\partial \gamma_1'}{\partial \omega}. $$
The function $\Psi$ has the following attribute:
$$ \Psi \sim \frac{1}{c_1 \varepsilon + c_2 \omega}. $$
We, therefore, have
$$ \Psi|_{\omega=0} \sim \frac{1}{\varepsilon}; \ \left. \frac{\partial \Psi}{\partial \omega} \right|_{\omega=0} \sim \frac{1}{\varepsilon^2}. $$
The graph $\gamma_1'$ has second order pole in $\varepsilon$ and differentiation in $\omega$ does not increase the singularity in $\varepsilon$:
$$ \gamma_1'|_{\omega=0} \sim \frac{1}{\varepsilon^2}; \ \left. \frac{\partial \gamma_1'}{\partial \omega} \right|_{\omega=0} \sim \frac{1}{\varepsilon^2}. $$
Therefore, to compute $\gamma_1$ with desired accuracy we need to calculate 4 terms in $\varepsilon$ for $\gamma_1'|_{\omega=0}$ and 3 terms for $\left. \frac{\partial \gamma_1'}{\partial \omega} \right|_{\omega=0}$.
Two button vertexes of the diagram $\gamma_1'$ are unique, and two top ones are not unique.
Let us consider the following combination:
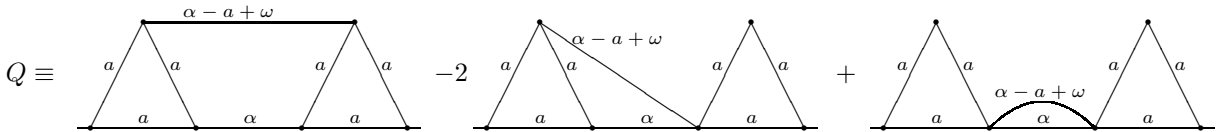
\begin{center}
\begin{picture}(500,50)
\put(3,23){$Q \equiv$}
\put(30,5){\line(1,0){130}}
\put(35,5){\circle*{2}}
\put(75,5){\circle*{2}}
\put(115,5){\circle*{2}}
\put(155,5){\circle*{2}}
\put(35,5){\line(1,2){20}}
\put(135,45){\line(1,-2){20}}
\put(55,45){\circle*{2}}
\put(135,45){\circle*{2}}
\put(55,45){\line(1,-2){20}}
\put(115,5){\line(1,2){20}}
\put(55,45){\line(1,0){80}}
\put(93,7){\scriptsize $\alpha$}
\put(53,7){\scriptsize $a$}
\put(133,7){\scriptsize $a$}
\put(40,27){\scriptsize $a$}
\put(120,27){\scriptsize $a$}
\put(65,27){\scriptsize $a$}
\put(145,27){\scriptsize $a$}
\put(70,47){\scriptsize $\alpha-a+\omega$}
\put(165,23){$-2$}
\put(180,5){\line(1,0){130}}
\put(185,5){\circle*{2}}
\put(225,5){\circle*{2}}
\put(265,5){\circle*{2}}
\put(305,5){\circle*{2}}
\put(185,5){\line(1,2){20}}
\put(285,45){\line(1,-2){20}}
\put(205,45){\circle*{2}}
\put(285,45){\circle*{2}}
\put(205,45){\line(1,-2){20}}
\put(265,5){\line(1,2){20}}
\put(205,45){\line(3,-2){60}}
\put(243,7){\scriptsize $\alpha$}
\put(203,7){\scriptsize $a$}
\put(283,7){\scriptsize $a$}
\put(190,27){\scriptsize $a$}
\put(270,27){\scriptsize $a$}
\put(215,27){\scriptsize $a$}
\put(295,27){\scriptsize $a$}
\put(217,37){\scriptsize $\alpha-a+\omega$}
\put(317,23){$+$}
\put(330,5){\line(1,0){130}}
\put(335,5){\circle*{2}}
\put(375,5){\circle*{2}}
\put(415,5){\circle*{2}}
\put(455,5){\circle*{2}}
\put(335,5){\line(1,2){20}}
\put(435,45){\line(1,-2){20}}
\put(355,45){\circle*{2}}
\put(435,45){\circle*{2}}
\put(355,45){\line(1,-2){20}}
\put(415,5){\line(1,2){20}}
\put(353,7){\scriptsize $a$}
\put(433,7){\scriptsize $a$}
\put(340,27){\scriptsize $a$}
\put(420,27){\scriptsize $a$}
\put(365,27){\scriptsize $a$}
\put(445,27){\scriptsize $a$}
\put(393,7){\scriptsize $\alpha$}
\qbezier(375,5)(395,25)(415,5)
\put(377,17){\scriptsize $\alpha-a+\omega$}
\end{picture}
\end{center}
All these graphs have second order poles in $\varepsilon$, whereas the combination $Q$ is finite. The second and third graphs in $Q$ are calculated explicitly.
The index of moved line is
$$\alpha-a+\omega=\left(\frac{1}{2}+\frac{3\eta_1}{4}\right)\varepsilon+ {\cal O}(\varepsilon^2) + \omega \equiv c_1 \varepsilon + c_2 \omega + ...$$
If we consider the case with $c_1=c_2=0$ then we obtain $Q = 0$. Therefore, in the linear approximation it holds:
$$ Q = \widetilde{C} (c_1 \varepsilon + c_2 \omega) + ... $$
Now, our task is to find the coefficient $\widetilde{C}$. The value $Q$ is finite, therefore, it does not depend on the way of regularization. We choose the regularization such as all the diagrams are calculated explicitly:
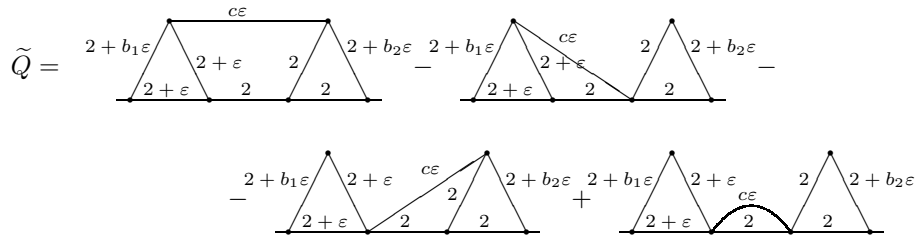
\begin{center}
\begin{picture}(335,90)
\put(0,65){$\widetilde{Q}=$}
\put(40,55){\line(1,0){100}}
\put(45,55){\circle*{2}}
\put(75,55){\circle*{2}}
\put(105,55){\circle*{2}}
\put(135,55){\circle*{2}}
\put(45,55){\line(1,2){15}}
\put(120,85){\line(1,-2){15}}
\put(60,85){\circle*{2}}
\put(120,85){\circle*{2}}
\put(60,85){\line(1,-2){15}}
\put(105,55){\line(1,2){15}}
\put(60,85){\line(1,0){60}}
\put(50,57){\scriptsize $2+\varepsilon$}
\put(70,67){\scriptsize $2+\varepsilon$}
\put(87,57){\scriptsize $2$}
\put(117,57){\scriptsize $2$}
\put(105,67){\scriptsize $2$}
\put(28,73){\scriptsize $2+b_1 \varepsilon$}
\put(127,73){\scriptsize $2+b_2 \varepsilon$}
\put(82,87){\scriptsize $c \varepsilon$}
\put(152,65){$-$}
\put(158,73){\scriptsize $2+b_1 \varepsilon$}
\put(170,55){\line(1,0){100}}
\put(175,55){\circle*{2}}
\put(205,55){\circle*{2}}
\put(235,55){\circle*{2}}
\put(265,55){\circle*{2}}
\put(175,55){\line(1,2){15}}
\put(250,85){\line(1,-2){15}}
\put(190,85){\circle*{2}}
\put(250,85){\circle*{2}}
\put(190,85){\line(1,-2){15}}
\put(235,55){\line(1,2){15}}
\put(190,85){\line(3,-2){45}}
\put(180,57){\scriptsize $2+\varepsilon$}
\put(200,67){\scriptsize $2+\varepsilon$}
\put(217,57){\scriptsize $2$}
\put(247,57){\scriptsize $2$}
\put(238,73){\scriptsize $2$}
\put(257,73){\scriptsize $2+b_2 \varepsilon$}
\put(207,77){\scriptsize $c \varepsilon$}
\put(282,65){$-$}
\put(82,15){$-$}
\put(88,23){\scriptsize $2+b_1 \varepsilon$}
\put(100,5){\line(1,0){100}}
\put(105,5){\circle*{2}}
\put(135,5){\circle*{2}}
\put(165,5){\circle*{2}}
\put(195,5){\circle*{2}}
\put(105,5){\line(1,2){15}}
\put(180,35){\line(1,-2){15}}
\put(120,35){\circle*{2}}
\put(180,35){\circle*{2}}
\put(120,35){\line(1,-2){15}}
\put(165,5){\line(1,2){15}}
\put(135,5){\line(3,2){45}}
\put(110,7){\scriptsize $2+\varepsilon$}
\put(127,23){\scriptsize $2+\varepsilon$}
\put(147,7){\scriptsize $2$}
\put(177,7){\scriptsize $2$}
\put(165,17){\scriptsize $2$}
\put(187,23){\scriptsize $2+b_2 \varepsilon$}
\put(155,27){\scriptsize $c \varepsilon$}
\put(212,15){$+$}
\put(218,23){\scriptsize $2+b_1 \varepsilon$}
\put(230,5){\line(1,0){100}}
\put(235,5){\circle*{2}}
\put(265,5){\circle*{2}}
\put(295,5){\circle*{2}}
\put(325,5){\circle*{2}}
\put(235,5){\line(1,2){15}}
\put(310,35){\line(1,-2){15}}
\put(250,35){\circle*{2}}
\put(310,35){\circle*{2}}
\put(250,35){\line(1,-2){15}}
\put(295,5){\line(1,2){15}}
\qbezier(265,5)(280,25)(295,5)
\put(240,7){\scriptsize $2+\varepsilon$}
\put(257,23){\scriptsize $2+\varepsilon$}
\put(277,7){\scriptsize $2$}
\put(307,7){\scriptsize $2$}
\put(298,23){\scriptsize $2$}
\put(317,23){\scriptsize $2+b_2 \varepsilon$}
\put(275,17){\scriptsize $c \varepsilon$}
\end{picture}
\end{center}
In the linear approximation we have $\widetilde{Q}=\widetilde{C}c\varepsilon + ...$ with the same coefficient $\widetilde{C}$.
We use the fact that the graph of the following type is calculated explicitly in arbitrary dimension:
\begin{center}
\begin{picture}(300,60)
\put(0,23){$G_n(m_1,...,m_n;\ \beta_1,...,\beta_{n-1}) \ \equiv $}
\put(150,10){\line(1,0){90}}
\put(155,10){\circle*{2}}
\put(195,10){\circle*{2}}
\put(235,10){\circle*{2}}
\put(160,30){\circle*{2}}
\put(175,45){\circle*{2}}
\put(195,50){\circle*{2}}
\put(230,30){\circle*{2}}
\put(155,10){\line(1,4){5}}
\put(160,30){\line(1,1){15}}
\put(175,45){\line(4,1){20}}
\put(161,30){\line(5,-3){34}}
\put(195,10){\line(5,3){34}}
\put(175,44){\line(3,-5){20}}
\put(195,10){\line(0,1){40}}
\put(230,30){\line(1,-4){5}}
\put(195,50){\line(4,-1){10}}
\put(230,30){\line(-1,1){10}}
\put(210,43){$...$}
\put(170,5){\scriptsize $m_1$}
\put(170,15){\scriptsize $m_2$}
\put(171,30){\scriptsize $m_3$}
\put(196,35){\scriptsize $m_4$}
\put(210,15){\scriptsize $m_{n-1}$}
\put(215,5){\scriptsize $m_n$}
\put(150,20){\scriptsize $\beta_1$}
\put(160,40){\scriptsize $\beta_2$}
\put(180,51){\scriptsize $\beta_3$}
\put(234,20){\scriptsize $\beta_{n-1}$}
\end{picture}
\end{center}
where $m_1, m_2, ..., m_n$ are positive integers, and $\beta_1, ..., \beta_{n-1}$ are arbitrary numbers.
For calculation of this diagram we repeatedly apply the formula of integration by parts: \\
\begin{picture}(420,40)
\put(10,30){\line(1,0){40}}
\put(30,10){\line(0,1){20}}
\put(10,30){\circle*{2}}
\put(30,30){\circle*{2}}
\put(50,30){\circle*{2}}
\put(30,10){\circle*{2}}
\put(31,19){\scriptsize $z_1$}
\put(17,32){\scriptsize $z_2$}
\put(37,32){\scriptsize $z_3$}
\put(55,20){$=\frac{1}{d-2z_1-z_2-z_3} \left\{ z_2 \left[ \frac{}{} \right. \right.$}
\put(145,30){\line(1,0){40}}
\put(165,10){\line(0,1){20}}
\put(145,30){\circle*{2}}
\put(165,30){\circle*{2}}
\put(185,30){\circle*{2}}
\put(165,10){\circle*{2}}
\put(166,19){\scriptsize $z_1-1$}
\put(144,32){\scriptsize $z_2+1$}
\put(172,32){\scriptsize $z_3$}
\put(195,20){$-$}
\put(210,30){\line(1,0){40}}
\put(230,10){\line(0,1){20}}
\put(210,30){\circle*{2}}
\put(230,30){\circle*{2}}
\put(250,30){\circle*{2}}
\put(230,10){\circle*{2}}
\put(231,19){\scriptsize $z_1$}
\put(209,32){\scriptsize $z_2+1$}
\put(237,32){\scriptsize $z_3$}
\put(210,30){\line(1,-1){20}}
\put(209,16){\scriptsize $-1$}
\put(252,20){$\left. \frac{}{} \right] + z_3 \left[ \frac{}{} \right. $}
\put(290,30){\line(1,0){40}}
\put(310,10){\line(0,1){20}}
\put(290,30){\circle*{2}}
\put(310,30){\circle*{2}}
\put(330,30){\circle*{2}}
\put(310,10){\circle*{2}}
\put(311,19){\scriptsize $z_1-1$}
\put(297,32){\scriptsize $z_2$}
\put(309,32){\scriptsize $z_3+1$}
\put(340,20){$-$}
\put(355,30){\line(1,0){40}}
\put(375,10){\line(0,1){20}}
\put(355,30){\circle*{2}}
\put(375,30){\circle*{2}}
\put(395,30){\circle*{2}}
\put(375,10){\circle*{2}}
\put(366,19){\scriptsize $z_1$}
\put(362,32){\scriptsize $z_2$}
\put(374,32){\scriptsize $z_3+1$}
\put(375,10){\line(1,1){20}}
\put(386,16){\scriptsize $-1$}
\put(401,20){$\left. \left. \frac{}{} \right] \right\}.$}

\end{picture}

After computations we obtain
$$ \widetilde{Q} = \left(-\frac{1}{2} + \zeta(3) \right) c \varepsilon + {\cal O}(\varepsilon^2) $$
where $\zeta(z)$ is the Riemann's zeta function.
It, therefore, holds $\widetilde{C}=-\frac{1}{2} + \zeta(3)$, and we receive the following result:
\begin{align}
& \gamma_1|_{\omega=0} = \frac{8\pi^{18}}{(2+3\eta_1)^3 \varepsilon^3} + \frac{3 \pi^{18}((2+3\eta_1)(3\eta_1-26+16\tau)-24\eta_2)}{(2+3\eta_1)^4 \varepsilon^2} + \nonumber\\
& + \frac{\pi^{18}}{4(2+3\eta_1)^5 \varepsilon} [16(36 \tau^2 -\pi^2 )(2 + 3 \eta_1)^2 + 72 \tau (2 + 3 \eta_1) (9\eta_1^2-72\eta_1-52-24\eta_2) + \nonumber \\
& + 3 (81 \eta_1^4 -1044 \eta_1^3 + 24 \eta_1^2 (140 - 9 \eta_2) + 16 \eta_1 (373 + 171 \eta_2 - 18 \eta_3) + 48 (45 + 40 \eta_2 + 12 \eta_2^2 - 4 \eta_3))] + \nonumber \\
& + \frac{\pi^{18}}{32(2+3\eta_1)^6} [( 9216 \tau^3 -47656 + 17076 \eta_1 - 2910 \eta_1^2 - 33 \eta_1^3)(2 + 3 \eta_1)^3  - \nonumber \\
& -24 (2+3\eta_1)^2 (5116 - 804 \eta_1 + 27 \eta_1^2) \eta_2 + 7776 (-18 + \eta_1) (2 + 3 \eta_1) \eta_2^2 - 69120 \eta_2^3 + \nonumber \\
& + 48(36 \tau^2 - \pi^2)(2 + 3 \eta_1)^2 (9\eta_1^2-72\eta_1-52-24\eta_2) + \nonumber \\
& +  144 \tau (2 + 3 \eta_1) (81 \eta_1^4 -1044 \eta_1^3 +
        24 \eta_1^2 (140 - 9 \eta_2) + 16 \eta_1 (373 + 171 \eta_2 - 18 \eta_3) + 48 (45 + 40 \eta_2 + 12 \eta_2^2 - 4 \eta_3)) - \nonumber \\
& - 768 \tau \pi^2 (2 + 3 \eta_1)^2 -576 (2 + 3 \eta_1) (-80 - 114 \eta_1 + 9 \eta_1^2 - 48 \eta_2) \eta_3 - 2304 ( 4 - 12 \eta_1 - 9 \eta_1^2) \eta_4 + \nonumber \\
& + 16 (2 + 3 \eta_1)^3 (32 + 36 \eta_1 + 60 \eta_1^2 + 21 \eta_1^3) \zeta(3)] +{\cal O}(\varepsilon), \nonumber
\end{align}
\begin{align}
& \left. \frac{\partial \gamma_1}{\partial \omega} \right|_{\omega=0} = -\frac{32 \pi^{18}}{(2+3 \eta_1)^4 \varepsilon^4} - \frac{8\pi^{18}((2+3\eta_1)(3\eta_1-40+24\tau) -48 \eta_2)}{(2+3 \eta_1)^5\varepsilon^3} - \nonumber\\
& - \frac{\pi^{18}}{(2+3\eta_1)^6\varepsilon^2} [16 (36 \tau^2 - \pi^2) (2 + 3 \eta_1)^2  + 48 \tau (2 + 3 \eta_1) (9 \eta_1^2 -114 \eta_1 -80 -48 \eta_2) + \nonumber\\
& +297 \eta_1^4 - 1728 \eta_1^3 +216(59 - 3 \eta_2) \eta_1^2 +32 (614 + 351 \eta_2 - 36 \eta_3) \eta_1 +48 (143 + 162 \eta_2 + 60 \eta_2^2 - 16 \eta_3) ] + \nonumber\\
& + \frac{\pi^{18}}{2(2+3\eta_1)^7\varepsilon} [(13092 -2516\eta_1 +649\eta_1^2 -78\eta_1^3 -2304 \tau^3)(2+3\eta_1)^3 +8 (36 \tau^2 -\pi^2) (2 + 3 \eta_1)^2 (80 +114 \eta_1 -  9 \eta_1^2 + 48 \eta_2) + \nonumber\\
& + 4 (2 + 3 \eta_1)^2 (10532 - 1128 \eta_1 + 99 \eta_1^2) \eta_2 +864 (136 + 198 \eta_1 - 9 \eta_1^2) \eta_2^2 +34560 \eta_2^3 + \nonumber\\
& + 12 \tau (2 + 3 \eta_1) (-297 \eta_1^4 +1728 \eta_1^3 + 216 \eta_1^2 (-59 + 3 \eta_2) -32 \eta_1 (614 + 351 \eta_2 - 36 \eta_3) -
48 (143 + 162 \eta_2 + 60 \eta_2^2- 16 \eta_3)) + \nonumber\\
& + 192 \tau \pi^2 (2 + 3 \eta_1)^2 +432 (2 + 3 \eta_1)^2(\eta_1 -18) \eta_3 -11520 (2 + 3\eta_1) \eta_2 \eta_3 + 768 (2 + 3 \eta_1)^2 \eta_4 + \nonumber\\
& + 8 (2 + 3 \eta_1)^3 (-12 + 4 \eta_1 + 5 \eta_1^2 + 6 \eta_1^3) \zeta(3) ] + {\cal O}(1) \nonumber
\end{align}
where $\tau = \gamma_E + \ln \pi$, and $\gamma_E$ is the Euler's constant.

Let us consider the diagram $\gamma_2$. Similarly, we make the function $\Psi$:
$$ \gamma_2 = \Psi \gamma_2' $$
and we receive the following graph $\gamma_2'$:
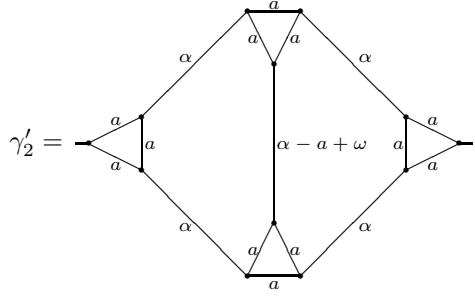
\begin{center}
\begin{picture}(175,110)
\put(0,53){$\gamma_2'=$}
\put(25,55){\line(1,0){5}}
\put(30,55){\circle*{2}}
\put(50,45){\circle*{2}}
\put(50,65){\circle*{2}}
\put(30,55){\line(2,1){20}}
\put(30,55){\line(2,-1){20}}
\put(50,45){\line(0,1){20}}
\put(50,45){\line(1,-1){40}}
\put(50,65){\line(1,1){40}}
\put(90,5){\circle*{2}}
\put(110,5){\circle*{2}}
\put(90,105){\circle*{2}}
\put(110,105){\circle*{2}}
\put(100,25){\circle*{2}}
\put(100,85){\circle*{2}}
\put(90,5){\line(1,0){20}}
\put(90,105){\line(1,0){20}}
\put(90,5){\line(1,2){10}}
\put(90,105){\line(1,-2){10}}
\put(100,25){\line(1,-2){10}}
\put(100,85){\line(1,2){10}}
\put(100,25){\line(0,1){60}}
\put(110,5){\line(1,1){40}}
\put(110,105){\line(1,-1){40}}
\put(150,45){\circle*{2}}
\put(150,65){\circle*{2}}
\put(170,55){\circle*{2}}
\put(150,45){\line(2,1){20}}
\put(150,65){\line(2,-1){20}}
\put(150,45){\line(0,1){20}}
\put(170,55){\line(1,0){5}}
\put(38,45){\scriptsize $a$}
\put(38,62){\scriptsize $a$}
\put(158,62){\scriptsize $a$}
\put(158,45){\scriptsize $a$}
\put(51,53){\scriptsize $a$}
\put(145,53){\scriptsize $a$}
\put(98,0){\scriptsize $a$}
\put(90,13){\scriptsize $a$}
\put(106,13){\scriptsize $a$}
\put(98,106){\scriptsize $a$}
\put(90,93){\scriptsize $a$}
\put(106,93){\scriptsize $a$}
\put(101,53){\scriptsize $\alpha-a+\omega$}
\put(64,22){\scriptsize $\alpha$}
\put(132,22){\scriptsize $\alpha$}
\put(64,86){\scriptsize $\alpha$}
\put(132,86){\scriptsize $\alpha$}
\end{picture}
\end{center}
To obtain $\eta_4$ we need to compute three terms in $\varepsilon$ for $\gamma_2'|_{\omega=0}$ and two terms for $\left. \frac{\partial \gamma_2'}{\partial \omega} \right|_{\omega=0}$. After integrations of unique vertexes we receive:
\begin{center}
\begin{picture}(390,110)
\put(0,53){$\gamma_2'= \pi^{2d} H(a,a,\alpha)^2 H(\alpha, a', a+\alpha')^2 $}
\put(175,55){\line(1,0){5}}
\put(180,55){\circle*{2}}
\put(230,5){\circle*{2}}
\put(230,105){\circle*{2}}
\put(280,30){\circle*{2}}
\put(280,80){\circle*{2}}
\put(330,5){\circle*{2}}
\put(330,105){\circle*{2}}
\put(380,55){\circle*{2}}
\put(180,55){\line(1,1){50}}
\put(180,55){\line(1,-1){50}}
\put(230,5){\line(0,1){100}}
\put(230,5){\line(1,0){100}}
\put(230,5){\line(2,1){50}}
\put(230,105){\line(1,0){100}}
\put(230,105){\line(2,-1){50}}
\put(280,30){\line(2,-1){50}}
\put(280,80){\line(2,1){50}}
\put(280,30){\line(0,1){50}}
\put(330,5){\line(0,1){100}}
\put(330,5){\line(1,1){50}}
\put(330,105){\line(1,-1){50}}
\put(380,55){\line(1,0){5}}
\put(281,53){\scriptsize $\alpha-a+\omega$}
\put(231,53){\scriptsize $\alpha-a$}
\put(331,53){\scriptsize $\alpha-a$}
\put(278,0){\scriptsize $a$}
\put(278,106){\scriptsize $a$}
\put(254,21){\scriptsize $a$}
\put(302,21){\scriptsize $a$}
\put(254,85){\scriptsize $a$}
\put(302,85){\scriptsize $a$}
\put(200,25){\scriptsize $a$}
\put(200,80){\scriptsize $a$}
\put(356,25){\scriptsize $a$}
\put(356,80){\scriptsize $a$}
\end{picture}
\end{center}
We consider the following combination:
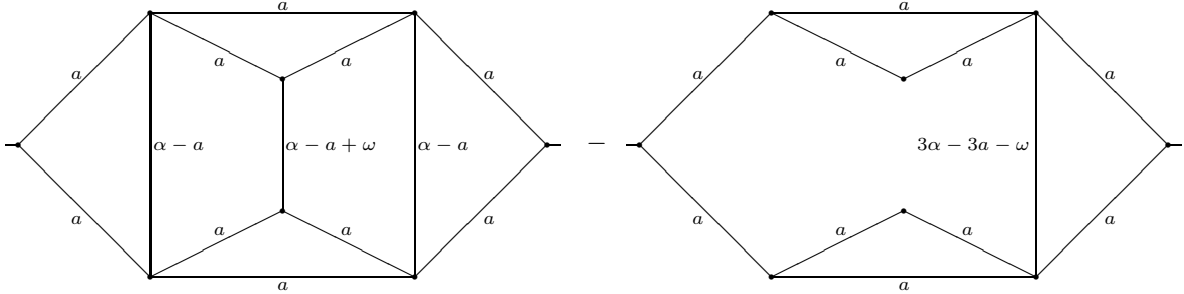
\begin{center}
\begin{picture}(450,110)
\put(0,55){\line(1,0){5}}
\put(5,55){\circle*{2}}
\put(55,5){\circle*{2}}
\put(55,105){\circle*{2}}
\put(105,30){\circle*{2}}
\put(105,80){\circle*{2}}
\put(155,5){\circle*{2}}
\put(155,105){\circle*{2}}
\put(205,55){\circle*{2}}
\put(5,55){\line(1,1){50}}
\put(5,55){\line(1,-1){50}}
\put(55,5){\line(1,0){100}}
\put(55,5){\line(2,1){50}}
\put(55,105){\line(1,0){100}}
\put(55,105){\line(2,-1){50}}
\put(105,30){\line(2,-1){50}}
\put(105,80){\line(2,1){50}}
\put(55,5){\line(0,1){100}}
\put(155,5){\line(0,1){100}}
\put(105,30){\line(0,1){50}}
\put(155,5){\line(1,1){50}}
\put(155,105){\line(1,-1){50}}
\put(205,55){\line(1,0){5}}
\put(56,53){\scriptsize $\alpha-a$}
\put(106,53){\scriptsize $\alpha-a+\omega$}
\put(156,53){\scriptsize $\alpha-a$}
\put(103,0){\scriptsize $a$}
\put(103,106){\scriptsize $a$}
\put(79,21){\scriptsize $a$}
\put(127,21){\scriptsize $a$}
\put(79,85){\scriptsize $a$}
\put(127,85){\scriptsize $a$}
\put(25,25){\scriptsize $a$}
\put(25,80){\scriptsize $a$}
\put(181,25){\scriptsize $a$}
\put(181,80){\scriptsize $a$}
\put(220,53){$-$}
\put(235,55){\line(1,0){5}}
\put(240,55){\circle*{2}}
\put(290,5){\circle*{2}}
\put(290,105){\circle*{2}}
\put(340,30){\circle*{2}}
\put(340,80){\circle*{2}}
\put(390,5){\circle*{2}}
\put(390,105){\circle*{2}}
\put(440,55){\circle*{2}}
\put(240,55){\line(1,1){50}}
\put(240,55){\line(1,-1){50}}
\put(290,5){\line(1,0){100}}
\put(290,5){\line(2,1){50}}
\put(290,105){\line(1,0){100}}
\put(290,105){\line(2,-1){50}}
\put(340,30){\line(2,-1){50}}
\put(340,80){\line(2,1){50}}
\put(390,5){\line(0,1){100}}
\put(390,5){\line(1,1){50}}
\put(390,105){\line(1,-1){50}}
\put(440,55){\line(1,0){5}}
\put(345,53){\scriptsize $3\alpha-3a-\omega$}
\put(338,0){\scriptsize $a$}
\put(338,106){\scriptsize $a$}
\put(314,21){\scriptsize $a$}
\put(362,21){\scriptsize $a$}
\put(314,85){\scriptsize $a$}
\put(362,85){\scriptsize $a$}
\put(260,25){\scriptsize $a$}
\put(260,80){\scriptsize $a$}
\put(416,25){\scriptsize $a$}
\put(416,80){\scriptsize $a$}
\end{picture}
\end{center}
This combination is finite when $\omega=0$, and it has first order pole in $\varepsilon$ at arbitrary $\omega$. We introduce auxiliary diagram
\begin{center}
\begin{picture}(220,90)
\put(0,43){$\xi \equiv $}
\put(20,45){\line(1,0){5}}
\put(25,45){\circle*{2}}
\put(185,45){\circle*{2}}
\put(105,5){\circle*{2}}
\put(105,85){\circle*{2}}
\put(25,45){\line(2,1){80}}
\put(25,45){\line(2,-1){80}}
\put(105,5){\line(2,1){80}}
\put(105,85){\line(2,-1){80}}
\put(105,5){\line(0,1){80}}
\put(185,45){\line(1,0){5}}
\put(10,75){\scriptsize $\frac{d}{2}-1+c_1\varepsilon+\widetilde{c_1}\varepsilon^2$}
\put(10,10){\scriptsize $\frac{d}{2}-1+c_4\varepsilon+\widetilde{c_4}\varepsilon^2$}
\put(130,75){\scriptsize $\frac{d}{2}-1+c_2\varepsilon+\widetilde{c_2}\varepsilon^2$}
\put(130,10){\scriptsize $\frac{d}{2}-1+c_3\varepsilon+\widetilde{c_3}\varepsilon^2$}
\put(106,43){\scriptsize $c_5 \varepsilon+\widetilde{c_5}\varepsilon^2+c_6 \omega$}
\put(209,20){.}
\end{picture}
\end{center}
It is finite at $\varepsilon=\omega=0$.
Computations of $\gamma_2$ with necessary accuracy are reduced to calculation of $\xi$ up to the terms of order $\varepsilon^2$ and $\varepsilon \omega$.
We do calculations by using the methods described in [1, 6---9]. We obtain:
\begin{align}
& \xi =\pi^6 + \pi^6 \left[-4+2\tau +\left( \frac{5}{3}-2\zeta(3) \right) c_5 \right] \varepsilon + \pi^6 \left(\frac{5}{3}-2\zeta(3)\right) c_6 \omega + \nonumber\\
& + \pi^6 \left[10 + 2 \tau (\tau -4) - \frac{\pi^2}{6} + c_1 + c_1^2 +c_2+c_2^2 +c_1c_2 + c_3+c_3^2+c_4+c_4^2+c_3c_4+ \right. \nonumber\\
& \left. + \left(\frac{10 \tau}{3} + \frac{\pi^4}{30} -\frac{58}{9} + \left( \frac{19}{3} - 4\tau \right) \zeta(3) + c_1+c_2+c_3+c_4 \right) c_5 +\left( \frac{5}{3}-2\zeta(3) \right) \widetilde{c_5} + \left( \frac{7}{3}-\zeta(3) \right) c_5^2 \right] \varepsilon^2 + \nonumber\\
& + \pi^6 \left[ \frac{10 \tau}{3} + \frac{\pi^4}{30} -\frac{58}{9} +\left(\frac{19}{3}-4\tau\right) \zeta(3) + c_1+c_2+c_3+c_4 + \left(\frac{14}{3}-2\zeta(3) \right) c_5 \right] c_6 \varepsilon \omega +... \nonumber
\end{align}
The result for $\gamma_2$ is the following:
\begin{align}
& \gamma_2 |_{\omega=0} = \frac{32 \pi^{36}}{(2+3 \eta_1)^5 \varepsilon^5} + \frac{4 \pi^{36} [(45\eta_1-142)(2 + 3 \eta_1) -120 \eta_2 + 96 (2 + 3 \eta_1) \tau -12(2 + 3 \eta_1)^2 \zeta(3)]}{(2+3 \eta_1)^6 \varepsilon^4} + \nonumber\\
& + \frac{\pi^{36}}{30 (2+3 \eta_1)^7 \varepsilon^3} [960(72\tau^2-\pi^2) (2+3 \eta_1)^2 + 24(\pi^4 -720\tau \zeta(3))(2+3 \eta_1)^3  \nonumber\\
& -1440 \tau (2+3 \eta_1) ((45 \eta_1-142)(2 + 3 \eta_1)+120 \eta_2) +5 (2+3 \eta_1)^2 (31372 - 19380 \eta_1 + 3555 \eta_1^2) \nonumber\\
& -7200(2+3 \eta_1)(9\eta_1-37)\eta_2 +129600\eta_2^2 -14400(2+3 \eta_1)\eta_3 -480 (2+3 \eta_1)^2 ((2+3 \eta_1)(9\eta_1-56)-36\eta_2)\zeta(3) ] + {\cal O} \left( \frac{1}{\varepsilon^2} \right), \nonumber
\end{align}
\begin{align}
& \left. \frac{\partial \gamma_2}{\partial \omega} \right|_{\omega=0} = -\frac{128 \pi^{36}}{(2+3 \eta_1)^6 \varepsilon^6} + \frac{64 \pi^{36} [(113-24\eta_1)(2 + 3 \eta_1) +108 \eta_2 -72 (2 + 3 \eta_1) \tau +6(2 + 3 \eta_1)^2 \zeta(3)]}{3(2+3 \eta_1)^7 \varepsilon^5} + \nonumber \\
& + \frac{2 \pi^{36}}{45 (2+3 \eta_1)^8 \varepsilon^4} [-2880(72\tau^2-\pi^2)(2+3 \eta_1)^2 -48\pi^4 (2+3 \eta_1)^3 -5(2+3 \eta_1)^2(105500 - 44844 \eta_1 + 4995 \eta_1^2) + \nonumber\\
& +2880 (2 + 3 \eta_1) (\eta_1 -347) \eta_2  -544320 \eta_2^2 + 51840 (2 + 3 \eta_1) \eta_3 +120(2 + 3 \eta_1)^2 ((2 + 3 \eta_1) (45 \eta_1-466 ) -360 \eta_2 ) \zeta(3) + \nonumber\\
& +5760\tau (2 + 3 \eta_1) (226 + 291 \eta_1 - 72 \eta_1^2 + 108 \eta_2 +
  6 (2 + 3 \eta_1)^2 \zeta(3))] +{\cal O} \left(\frac{1}{\varepsilon^3}\right). \nonumber
\end{align}

Let us consider the third order diagrams: $\gamma_{31}$, $\gamma_{32a}$, $\gamma_{32b}$ and $\gamma_{32c}$. The function $\Psi$ appears by the same way.
The rest graphs are the following:
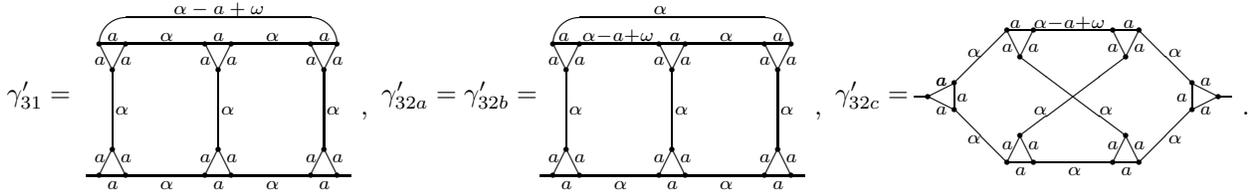
\begin{center}
\begin{picture}(135,70)
\put(0,33){$\gamma_{31}'=$}
\put(30,5){\line(1,0){100}}
\put(35,5){\circle*{2}}
\put(45,5){\circle*{2}}
\put(40,15){\circle*{2}}
\put(75,5){\circle*{2}}
\put(85,5){\circle*{2}}
\put(80,15){\circle*{2}}
\put(115,5){\circle*{2}}
\put(125,5){\circle*{2}}
\put(120,15){\circle*{2}}
\put(35,5){\line(1,2){5}}
\put(75,5){\line(1,2){5}}
\put(115,5){\line(1,2){5}}
\put(40,15){\line(1,-2){5}}
\put(80,15){\line(1,-2){5}}
\put(120,15){\line(1,-2){5}}
\put(35,55){\line(1,0){90}}
\put(35,55){\circle*{2}}
\put(45,55){\circle*{2}}
\put(40,45){\circle*{2}}
\put(75,55){\circle*{2}}
\put(85,55){\circle*{2}}
\put(80,45){\circle*{2}}
\put(115,55){\circle*{2}}
\put(125,55){\circle*{2}}
\put(120,45){\circle*{2}}
\put(35,55){\line(1,-2){5}}
\put(75,55){\line(1,-2){5}}
\put(115,55){\line(1,-2){5}}
\put(40,45){\line(1,2){5}}
\put(80,45){\line(1,2){5}}
\put(120,45){\line(1,2){5}}
\put(40,15){\line(0,1){30}}
\put(80,15){\line(0,1){30}}
\put(120,15){\line(0,1){30}}
\put(80,55){\oval(90,20)[t]}
\put(38,0){\scriptsize $a$}
\put(33,10){\scriptsize $a$}
\put(43,10){\scriptsize $a$}
\put(78,0){\scriptsize $a$}
\put(73,10){\scriptsize $a$}
\put(83,10){\scriptsize $a$}
\put(118,0){\scriptsize $a$}
\put(113,10){\scriptsize $a$}
\put(123,10){\scriptsize $a$}
\put(38,56){\scriptsize $a$}
\put(33,47){\scriptsize $a$}
\put(43,47){\scriptsize $a$}
\put(78,56){\scriptsize $a$}
\put(73,47){\scriptsize $a$}
\put(83,47){\scriptsize $a$}
\put(118,56){\scriptsize $a$}
\put(113,47){\scriptsize $a$}
\put(123,47){\scriptsize $a$}
\put(58,0){\scriptsize $\alpha$}
\put(98,0){\scriptsize $\alpha$}
\put(58,56){\scriptsize $\alpha$}
\put(98,56){\scriptsize $\alpha$}
\put(41,28){\scriptsize $\alpha$}
\put(81,28){\scriptsize $\alpha$}
\put(121,28){\scriptsize $\alpha$}
\put(63,66){\scriptsize $\alpha-a+\omega$}
\put(134,28){,}
\end{picture} \
\begin{picture}(165,70)
\put(0,33){$\gamma_{32a}'=\gamma_{32b}'=$}
\put(60,5){\line(1,0){100}}
\put(65,5){\circle*{2}}
\put(75,5){\circle*{2}}
\put(70,15){\circle*{2}}
\put(105,5){\circle*{2}}
\put(115,5){\circle*{2}}
\put(110,15){\circle*{2}}
\put(145,5){\circle*{2}}
\put(155,5){\circle*{2}}
\put(150,15){\circle*{2}}
\put(65,5){\line(1,2){5}}
\put(105,5){\line(1,2){5}}
\put(145,5){\line(1,2){5}}
\put(70,15){\line(1,-2){5}}
\put(110,15){\line(1,-2){5}}
\put(150,15){\line(1,-2){5}}
\put(65,55){\line(1,0){90}}
\put(65,55){\circle*{2}}
\put(75,55){\circle*{2}}
\put(70,45){\circle*{2}}
\put(105,55){\circle*{2}}
\put(115,55){\circle*{2}}
\put(110,45){\circle*{2}}
\put(145,55){\circle*{2}}
\put(155,55){\circle*{2}}
\put(150,45){\circle*{2}}
\put(65,55){\line(1,-2){5}}
\put(105,55){\line(1,-2){5}}
\put(145,55){\line(1,-2){5}}
\put(70,45){\line(1,2){5}}
\put(110,45){\line(1,2){5}}
\put(150,45){\line(1,2){5}}
\put(70,15){\line(0,1){30}}
\put(110,15){\line(0,1){30}}
\put(150,15){\line(0,1){30}}
\put(110,55){\oval(90,20)[t]}
\put(68,0){\scriptsize $a$}
\put(63,10){\scriptsize $a$}
\put(73,10){\scriptsize $a$}
\put(108,0){\scriptsize $a$}
\put(103,10){\scriptsize $a$}
\put(113,10){\scriptsize $a$}
\put(148,0){\scriptsize $a$}
\put(143,10){\scriptsize $a$}
\put(153,10){\scriptsize $a$}
\put(67,56){\scriptsize $a$}
\put(63,47){\scriptsize $a$}
\put(73,47){\scriptsize $a$}
\put(109,56){\scriptsize $a$}
\put(103,47){\scriptsize $a$}
\put(113,47){\scriptsize $a$}
\put(148,56){\scriptsize $a$}
\put(143,47){\scriptsize $a$}
\put(153,47){\scriptsize $a$}
\put(88,0){\scriptsize $\alpha$}
\put(128,0){\scriptsize $\alpha$}
\put(76,56){\scriptsize $\alpha{-}a{+}\omega$}
\put(128,56){\scriptsize $\alpha$}
\put(71,28){\scriptsize $\alpha$}
\put(111,28){\scriptsize $\alpha$}
\put(151,28){\scriptsize $\alpha$}
\put(103,66){\scriptsize $\alpha$}
\put(164,28){,}
\end{picture} \
\begin{picture}(155,70)
\put(0,33){$\gamma_{32c}'=$}
\put(30,35){\line(1,0){5}}
\put(35,35){\circle*{2}}
\put(35,35){\line(2,1){10}}
\put(35,35){\line(2,-1){10}}
\put(45,30){\line(0,1){10}}
\put(45,30){\circle*{2}}
\put(45,40){\circle*{2}}
\put(45,40){\line(1,1){20}}
\put(45,30){\line(1,-1){20}}
\put(65,10){\circle*{2}}
\put(65,60){\circle*{2}}
\put(75,10){\circle*{2}}
\put(75,60){\circle*{2}}
\put(70,20){\circle*{2}}
\put(70,50){\circle*{2}}
\put(105,10){\circle*{2}}
\put(105,60){\circle*{2}}
\put(115,10){\circle*{2}}
\put(115,60){\circle*{2}}
\put(110,20){\circle*{2}}
\put(110,50){\circle*{2}}
\put(65,10){\line(1,0){50}}
\put(65,60){\line(1,0){50}}
\put(65,10){\line(1,2){5}}
\put(65,60){\line(1,-2){5}}
\put(70,20){\line(1,-2){5}}
\put(70,50){\line(1,2){5}}
\put(105,10){\line(1,2){5}}
\put(105,60){\line(1,-2){5}}
\put(110,20){\line(1,-2){5}}
\put(110,50){\line(1,2){5}}
\put(115,10){\line(1,1){20}}
\put(115,60){\line(1,-1){20}}
\put(135,30){\circle*{2}}
\put(135,40){\circle*{2}}
\put(145,35){\circle*{2}}
\put(135,30){\line(0,1){10}}
\put(135,30){\line(2,1){10}}
\put(135,40){\line(2,-1){10}}
\put(145,35){\line(1,0){5}}
\put(70,20){\line(4,3){40}}
\put(70,50){\line(4,-3){40}}
\put(75,61){\scriptsize $\alpha{-}a{+}\omega$}
\put(68,5){\scriptsize $a$}
\put(63,15){\scriptsize $a$}
\put(73,15){\scriptsize $a$}
\put(108,5){\scriptsize $a$}
\put(103,15){\scriptsize $a$}
\put(113,15){\scriptsize $a$}
\put(66,61){\scriptsize $a$}
\put(63,52){\scriptsize $a$}
\put(73,52){\scriptsize $a$}
\put(109,61){\scriptsize $a$}
\put(103,52){\scriptsize $a$}
\put(113,52){\scriptsize $a$}
\put(38,39){\scriptsize $a$}
\put(38,28){\scriptsize $a$}
\put(138,39){\scriptsize $a$}
\put(138,28){\scriptsize $a$}
\put(38,39){\scriptsize $a$}
\put(46,33){\scriptsize $a$}
\put(129,33){\scriptsize $a$}
\put(50,50){\scriptsize $\alpha$}
\put(50,17){\scriptsize $\alpha$}
\put(126,50){\scriptsize $\alpha$}
\put(126,17){\scriptsize $\alpha$}
\put(75,28){\scriptsize $\alpha$}
\put(100,28){\scriptsize $\alpha$}
\put(88,5){\scriptsize $\alpha$}
\put(154,28){.}
\end{picture}
\end{center}
To obtain $\eta_4$ we need to find two terms in $\varepsilon$ for these graphs at $\omega=0$ and the main approximation in $\varepsilon$ at arbitrary $\omega$ up to linear term in $\omega$. Using analogical techniques we receive
\begin{align}
& \gamma_{31}|_{\omega=0} = \frac{128 \pi^{54}}{(2+3\eta_1)^7 \varepsilon^7} + \frac{16 \pi^{54}[(2+3\eta_1)(5(51\eta_1-122)+432\tau)-504\eta_2+32(2+3\eta_1)^2(2\zeta(3)-5\zeta(5))] }{(2+3\eta_1)^8 \varepsilon^6} +{\cal O} \left(\frac{1}{\varepsilon^5}\right), \nonumber
\end{align}
$$ \left. \frac{\partial \gamma_{31}}{\partial \omega} \right|_{\omega=0} = -\frac{512 \pi^{54}}{(2+3\eta_1)^8\varepsilon^8} - \frac{128 \pi^{54} [3(2+3\eta_1)((11\eta_1-36)+24\tau)-96\eta_2+4(2+3\eta_1)^2(2\zeta(3)-5\zeta(5))]}{(2+3\eta_1)^9\varepsilon^7} +{\cal O}\left(\frac{1}{\varepsilon^6}\right), $$
\begin{align}
& \gamma_{32a}|_{\omega=0} = \gamma_{32b}|_{\omega=0} = \gamma_{32c}|_{\omega=0} = \frac{128 (3\zeta(3)-1)\pi^{54}}{3(2+3 \eta_1)^7\varepsilon^7} - \nonumber\\
& - \frac{16\pi^{54}[(2+3 \eta_1)(6\pi^4-2160\tau(3\zeta(3)-1) -5(122 - 51 \eta_1)+15(706 - 63 \eta_1)\zeta(3))+2520(3\zeta(3)-1)\eta_2 ]}{45(2+3 \eta_1)^8\varepsilon^6} + {\cal O}\left(\frac{1}{\varepsilon^5}\right), \nonumber
\end{align}
\begin{align}
& \left. \frac{\partial \gamma_{32a}}{\partial \omega} \right|_{\omega=0} = \left. \frac{\partial \gamma_{32b}}{\partial \omega} \right|_{\omega=0} = -\frac{512 \pi^{54} (3\zeta(3)-1)}{3(2+3\eta_1)^8\varepsilon^8} + \nonumber\\
& + \frac{128 \pi^{54}[(2+3\eta_1)(3 \pi^4 - 1080 \tau (3\zeta(3)-1) +5 (96 \eta_1 -326 ) +15 (356 - 27 \eta_1) \zeta(3) ) +1440 (3\zeta(3)-1) \eta_2]}{45(2+3\eta_1)^9\varepsilon^7} +{\cal O}\left(\frac{1}{\varepsilon^6}\right), \nonumber
\end{align}
\begin{align}
& \left. \frac{\partial \gamma_{32c}}{\partial \omega} \right|_{\omega=0} =  -\frac{512 \pi^{54} (3\zeta(3)-1)}{3(2+3\eta_1)^8\varepsilon^8} + \nonumber\\
& + \frac{128 \pi^{54}[(2+3\eta_1)(3 \pi^4 - 1080 \tau (3\zeta(3)-1) +25 (21 \eta_1 -64 ) +15 (356 - 27 \eta_1) \zeta(3) ) +1440 (3\zeta(3)-1) \eta_2]}{45(2+3\eta_1)^9\varepsilon^7} +{\cal O}\left(\frac{1}{\varepsilon^6}\right). \nonumber
\end{align}

For the fourth order graphs $\gamma_{41}$ till $\gamma_{49}$ we should calculate the main approximation in $\varepsilon$ only.
After making the function $\Psi$ it is sufficient to calculate the main term in $\varepsilon$ of rest diagrams $\gamma_{4i}'$ at $\omega=0$ only. The singular contributions appear with the integration of vertex functions, these contributions are equal to
$$ Sing \left(\pi^d H(a,a,\alpha) H(\alpha,a+\alpha',a')\right) = \frac{2\pi^6}{(2+3\eta_1)\varepsilon}. $$
Every diagram $\gamma_{4i}'$ has 8 vertex functions.
$$ \gamma_{4i}' = \left(\frac{2\pi^6}{(2+3\eta_1)\varepsilon}\right)^8 \gamma_{4i}'' .$$
The graphs $\gamma_{4i}''$ do not have any singularities, and we can calculate them directly in the dimension $d=6$.
We obtain the following: \\
\begin{picture}(200,55)
\put(0,23){$\gamma_{41}''=$}
\put(30,25){\line(1,0){5}}
\put(35,25){\circle*{2}}
\put(55,45){\circle*{2}}
\put(55,5){\circle*{2}}
\put(85,45){\circle*{2}}
\put(85,5){\circle*{2}}
\put(105,25){\circle*{2}}
\put(105,25){\line(1,0){5}}
\put(35,25){\line(1,1){20}}
\put(35,25){\line(1,-1){20}}
\put(85,5){\line(1,1){20}}
\put(85,45){\line(1,-1){20}}
\put(55,5){\line(1,0){30}}
\put(55,45){\line(1,0){30}}
\put(55,5){\line(0,1){40}}
\put(85,5){\line(0,1){40}}
\put(45,35){\circle*{2}}
\put(95,15){\circle*{2}}
\put(37,32){\scriptsize $2$}
\put(47,42){\scriptsize $2$}
\put(41,10){\scriptsize $2$}
\put(56,23){\scriptsize $2$}
\put(80,23){\scriptsize $2$}
\put(68,0){\scriptsize $2$}
\put(68,46){\scriptsize $2$}
\put(90,5){\scriptsize $2$}
\put(100,15){\scriptsize $2$}
\put(96,35){\scriptsize $2$}
\put(115,23){$=\pi^{18}\left[-\frac{1}{3}+\zeta(3)\right],$}
\end{picture} \
\begin{picture}(230,55)
\put(0,23){$\gamma_{42}''=$}
\put(30,25){\line(1,0){5}}
\put(35,25){\circle*{2}}
\put(55,45){\circle*{2}}
\put(55,5){\circle*{2}}
\put(85,45){\circle*{2}}
\put(85,5){\circle*{2}}
\put(105,25){\circle*{2}}
\put(105,25){\line(1,0){5}}
\put(35,25){\line(1,1){20}}
\put(35,25){\line(1,-1){20}}
\put(85,5){\line(1,1){20}}
\put(85,45){\line(1,-1){20}}
\put(55,5){\line(1,0){30}}
\put(55,45){\line(1,0){30}}
\put(55,5){\line(3,4){30}}
\put(55,45){\line(3,-4){30}}
\put(45,35){\circle*{2}}
\put(95,15){\circle*{2}}
\put(37,32){\scriptsize $2$}
\put(47,42){\scriptsize $2$}
\put(41,10){\scriptsize $2$}
\put(56,33){\scriptsize $2$}
\put(80,33){\scriptsize $2$}
\put(68,0){\scriptsize $2$}
\put(68,46){\scriptsize $2$}
\put(90,5){\scriptsize $2$}
\put(100,15){\scriptsize $2$}
\put(96,35){\scriptsize $2$}
\put(115,23){$=\pi^{18}\left[-\frac{7}{3}\zeta(3)+\frac{10}{3}\zeta(5)\right],$}
\end{picture} \\
\begin{picture}(230,75)
\put(0,33){$\gamma_{43}''=$}
\put(30,35){\line(1,0){5}}
\put(35,35){\circle*{2}}
\put(55,55){\circle*{2}}
\put(55,15){\circle*{2}}
\put(85,55){\circle*{2}}
\put(85,15){\circle*{2}}
\put(105,35){\circle*{2}}
\put(105,35){\line(1,0){5}}
\put(35,35){\line(1,1){20}}
\put(35,35){\line(1,-1){20}}
\put(85,15){\line(1,1){20}}
\put(85,55){\line(1,-1){20}}
\put(55,15){\line(1,0){30}}
\put(55,55){\line(1,0){30}}
\put(55,15){\line(0,1){40}}
\put(85,15){\line(0,1){40}}
\put(55,35){\circle*{2}}
\put(95,25){\circle*{2}}
\put(41,46){\scriptsize $2$}
\put(41,20){\scriptsize $2$}
\put(56,23){\scriptsize $2$}
\put(56,43){\scriptsize $2$}
\put(80,33){\scriptsize $2$}
\put(68,10){\scriptsize $2$}
\put(68,56){\scriptsize $2$}
\put(90,15){\scriptsize $2$}
\put(100,25){\scriptsize $2$}
\put(96,45){\scriptsize $2$}
\put(115,33){$=\pi^{18}\left[-\frac{7}{3}\zeta(3)+\frac{10}{3}\zeta(5)\right],$}
\end{picture} \
\begin{picture}(250,75)
\put(0,33){$\gamma_{44}''=$}
\put(35,35){\line(1,0){5}}
\put(40,35){\circle*{2}}
\put(70,5){\circle*{2}}
\put(100,5){\circle*{2}}
\put(130,35){\circle*{2}}
\put(85,35){\circle*{2}}
\put(85,65){\circle*{2}}
\put(40,35){\line(1,-1){30}}
\put(100,5){\line(1,1){30}}
\put(40,35){\line(3,2){45}}
\put(85,65){\line(3,-2){45}}
\put(70,5){\line(1,0){30}}
\put(70,5){\line(1,2){15}}
\put(85,35){\line(1,-2){15}}
\put(85,35){\line(0,1){30}}
\put(130,35){\line(1,0){5}}
\put(64,51){\circle*{2}}
\put(85,5){\circle*{2}}
\put(76,0){\scriptsize $2$}
\put(91,0){\scriptsize $2$}
\put(72,18){\scriptsize $2$}
\put(94,18){\scriptsize $2$}
\put(51,15){\scriptsize $2$}
\put(115,15){\scriptsize $2$}
\put(49,45){\scriptsize $2$}
\put(67,57){\scriptsize $2$}
\put(110,50){\scriptsize $2$}
\put(86,45){\scriptsize $2$}
\put(140,33){$=\pi^{18}\left[-\frac{7}{3}\zeta(3)+\frac{10}{3}\zeta(5)\right],$}
\end{picture} \\
\begin{picture}(245,55)
\put(0,23){$\gamma_{45}''=$}
\put(30,25){\line(1,0){5}}
\put(35,25){\circle*{2}}
\put(55,45){\circle*{2}}
\put(55,5){\circle*{2}}
\put(85,45){\circle*{2}}
\put(85,5){\circle*{2}}
\put(105,25){\circle*{2}}
\put(105,25){\line(1,0){5}}
\put(35,25){\line(1,1){20}}
\put(35,25){\line(1,-1){20}}
\put(85,5){\line(1,1){20}}
\put(85,45){\line(1,-1){20}}
\put(55,5){\line(1,0){30}}
\put(55,45){\line(1,0){30}}
\put(55,5){\line(0,1){40}}
\put(85,5){\line(0,1){40}}
\put(45,35){\circle*{2}}
\put(95,35){\circle*{2}}
\put(37,32){\scriptsize $2$}
\put(47,42){\scriptsize $2$}
\put(41,10){\scriptsize $2$}
\put(56,23){\scriptsize $2$}
\put(80,23){\scriptsize $2$}
\put(68,0){\scriptsize $2$}
\put(68,46){\scriptsize $2$}
\put(90,42){\scriptsize $2$}
\put(100,32){\scriptsize $2$}
\put(96,10){\scriptsize $2$}
\put(115,23){$=\pi^{18}\left[\frac{1}{3}+\frac{10}{3}\zeta(3)-\frac{10}{3}\zeta(5)\right],$}
\end{picture} \
\begin{picture}(215,55)
\put(0,23){$\gamma_{46}''=$}
\put(30,25){\line(1,0){5}}
\put(35,25){\circle*{2}}
\put(55,45){\circle*{2}}
\put(55,5){\circle*{2}}
\put(85,45){\circle*{2}}
\put(85,5){\circle*{2}}
\put(105,25){\circle*{2}}
\put(105,25){\line(1,0){5}}
\put(35,25){\line(1,1){20}}
\put(35,25){\line(1,-1){20}}
\put(85,5){\line(1,1){20}}
\put(85,45){\line(1,-1){20}}
\put(55,5){\line(1,0){30}}
\put(55,45){\line(1,0){30}}
\put(55,5){\line(0,1){40}}
\put(85,5){\line(0,1){40}}
\put(55,25){\circle*{2}}
\put(85,25){\circle*{2}}
\put(41,36){\scriptsize $2$}
\put(41,10){\scriptsize $2$}
\put(56,13){\scriptsize $2$}
\put(56,33){\scriptsize $2$}
\put(80,33){\scriptsize $2$}
\put(80,13){\scriptsize $2$}
\put(68,0){\scriptsize $2$}
\put(68,46){\scriptsize $2$}
\put(96,10){\scriptsize $2$}
\put(96,35){\scriptsize $2$}
\put(115,23){$=\frac{10}{3}\pi^{18}[\zeta(3)-\zeta(5)],$}
\end{picture} \\
\begin{picture}(240,75)
\put(0,33){$\gamma_{47}''=$}
\put(35,35){\line(1,0){5}}
\put(40,35){\circle*{2}}
\put(70,5){\circle*{2}}
\put(100,5){\circle*{2}}
\put(130,35){\circle*{2}}
\put(85,35){\circle*{2}}
\put(85,65){\circle*{2}}
\put(40,35){\line(1,-1){30}}
\put(100,5){\line(1,1){30}}
\put(40,35){\line(3,2){45}}
\put(85,65){\line(3,-2){45}}
\put(70,5){\line(1,0){30}}
\put(70,5){\line(1,2){15}}
\put(85,35){\line(1,-2){15}}
\put(85,35){\line(0,1){30}}
\put(130,35){\line(1,0){5}}
\put(64,51){\circle*{2}}
\put(92,21){\circle*{2}}
\put(83,0){\scriptsize $2$}
\put(96,13){\scriptsize $2$}
\put(72,18){\scriptsize $2$}
\put(90,26){\scriptsize $2$}
\put(51,15){\scriptsize $2$}
\put(115,15){\scriptsize $2$}
\put(49,45){\scriptsize $2$}
\put(67,57){\scriptsize $2$}
\put(110,50){\scriptsize $2$}
\put(86,45){\scriptsize $2$}
\put(140,33){$=\frac{10}{3}\pi^{18}[\zeta(3)-\zeta(5)],$}
\end{picture} \
\begin{picture}(230,75)
\put(0,33){$\gamma_{48}''=$}
\put(30,35){\line(1,0){5}}
\put(35,35){\circle*{2}}
\put(55,55){\circle*{2}}
\put(55,15){\circle*{2}}
\put(85,55){\circle*{2}}
\put(85,15){\circle*{2}}
\put(105,35){\circle*{2}}
\put(105,35){\line(1,0){5}}
\put(35,35){\line(1,1){20}}
\put(35,35){\line(1,-1){20}}
\put(85,15){\line(1,1){20}}
\put(85,55){\line(1,-1){20}}
\put(55,15){\line(1,0){30}}
\put(55,55){\line(1,0){30}}
\put(55,15){\line(3,4){30}}
\put(55,55){\line(3,-4){30}}
\put(45,45){\circle*{2}}
\put(67,31){\circle*{2}}
\put(37,42){\scriptsize $2$}
\put(47,52){\scriptsize $2$}
\put(41,20){\scriptsize $2$}
\put(56,43){\scriptsize $2$}
\put(80,43){\scriptsize $2$}
\put(68,10){\scriptsize $2$}
\put(68,56){\scriptsize $2$}
\put(62,20){\scriptsize $2$}
\put(96,20){\scriptsize $2$}
\put(96,45){\scriptsize $2$}
\put(115,33){$=\frac{10}{3}\pi^{18}[\zeta(3)-\zeta(5)],$}
\end{picture} \\
\begin{picture}(220,55)
\put(0,23){$\gamma_{49}''=$}
\put(30,25){\line(1,0){5}}
\put(35,25){\circle*{2}}
\put(55,45){\circle*{2}}
\put(55,5){\circle*{2}}
\put(85,45){\circle*{2}}
\put(85,5){\circle*{2}}
\put(105,25){\circle*{2}}
\put(105,25){\line(1,0){5}}
\put(35,25){\line(1,1){20}}
\put(35,25){\line(1,-1){20}}
\put(85,5){\line(1,1){20}}
\put(85,45){\line(1,-1){20}}
\put(55,5){\line(1,0){30}}
\put(55,45){\line(1,0){30}}
\put(55,5){\line(3,4){30}}
\put(55,45){\line(3,-4){30}}
\put(73,21){\circle*{2}}
\put(67,21){\circle*{2}}
\put(41,35){\scriptsize $2$}
\put(79,13){\scriptsize $2$}
\put(41,10){\scriptsize $2$}
\put(56,33){\scriptsize $2$}
\put(80,33){\scriptsize $2$}
\put(67,0){\scriptsize $2$}
\put(67,46){\scriptsize $2$}
\put(62,10){\scriptsize $2$}
\put(96,10){\scriptsize $2$}
\put(96,35){\scriptsize $2$}
\put(115,23){$= \pi^{18}\left[-4\zeta(3)+5\zeta(5)\right].$}
\end{picture}

We use the notation: $\gamma_{4s}$ is the sum of all fourth order graphs with corresponding symmetry coefficients:
$$ \gamma_{4s} \equiv 3 \gamma_{41} + 3 \gamma_{42} + 6 \gamma_{43} + 6 \gamma_{44} + 3 \gamma_{45} + \frac{3}{2} \gamma_{46} + 3 \gamma_{47} + 3 \gamma_{48} + \frac{1}{2} \gamma_{49} .$$
We receive
$$ \gamma_{4s}'' = \pi^{18}\left[\zeta(3)+\frac{35}{2}\zeta(5)\right] $$
and
$$ \gamma_{4s}|_{\omega=0} = \frac{2^8 \pi^{72} (2\zeta(3)+35\zeta(5))}{(2+3\eta_1)^9 \varepsilon^9} + {\cal O} \left(\frac{1}{\varepsilon^8}\right), $$
$$ \left. \frac{\partial \gamma_{4s}}{\partial \omega} \right|_{\omega=0} = \frac{2^{10} \pi^{72} (2\zeta(3)+35\zeta(5))}{(2+3\eta_1)^{10} \varepsilon^{10}} + {\cal O} \left(\frac{1}{\varepsilon^9}\right). $$

\section{Results.}

After substituting the expressions for diagrams in (\ref{bootsys}) we obtain the result:
$$ \eta= \frac{2}{9}\varepsilon - \frac{172}{729}\varepsilon^2 +\left(\frac{16750}{3^{10}} - \frac{128 \zeta(3)}{3^5}\right)\varepsilon^3 +\left(-\frac{3883409}{3^{14}}-\frac{11456\zeta(3)}{3^9}+\frac{64\zeta(4)}{3^4}-\frac{1280\zeta(5)}{3^7}\right)\varepsilon^4 + {\cal O}(\varepsilon^5),\ d=6+2\varepsilon, $$
The three-loop approximation coincides with already known result for $\eta$ \cite{10,11}:
$$ \eta = -\frac{1}{9} \epsilon - \frac{43}{729} \epsilon^2 + \left(-\frac{8375}{2^2 3^{10}} + \frac{16 \zeta(3)}{3^5} \right) \epsilon^3 + {\cal O} (\epsilon^4), \ d=6-\epsilon. $$
The four-loop result for $\eta$ agrees with its numerical value obtained in \cite{12}:
$$ \eta = -0.1111 \epsilon - 0.0588 \epsilon^2 + 0.0436 \epsilon^3 - 0.081 \epsilon^4 + {\cal O} (\epsilon^5), \ d=6-\epsilon. $$
If we choose $d=6-\epsilon$, then our result will read:
$$ \eta = -\frac{1}{9} \epsilon - \frac{43}{729} \epsilon^2 + \left(-\frac{8375}{2^2 3^{10}} + \frac{16 \zeta(3)}{3^5} \right) \epsilon^3 + \left(-\frac{3883409}{2^4 3^{14}}-\frac{716\zeta(3)}{3^9}+\frac{4\zeta(4)}{3^4}-\frac{80\zeta(5)}{3^7}\right) \epsilon^4 + {\cal O} (\epsilon^5), $$
that is, $\eta_4=-\frac{3883409}{2^4 3^{14}}-\frac{716\zeta(3)}{3^9}+\frac{4\zeta(4)}{3^4}-\frac{80\zeta(5)}{3^7} =-0.07895... $
The numbers -0.081 and -0.07895 are very close.

In addition to the critical exponent $\eta$ we found the renorm-invariant combination of amplitudes $u$:
\begin{align}
& u=\frac{64}{27\pi^{18}} \varepsilon^3 + \frac{64 (217 - 162 \tau)}{729\pi^{18}} \varepsilon^4 +\frac{32 (4814 + 36 \tau (-217 + 81 \tau) + 81 \pi^2 -
   720 \zeta(3))}{3^7\pi^{18}} \varepsilon^5 +\nonumber\\
&+ \frac{32}{3^{15}\pi^{18}} [69182381 + 4271211 \pi^2 - 39366 (-3906 \tau^2 + 972 \tau^3 - 2 \pi^4 + \tau (4814 + 81 \pi^2 - 720 \zeta(3))) -\nonumber\\
&- 42655248 \zeta(3) - 12597120 \zeta(5)] \varepsilon^6 + {\cal O}(\varepsilon^7). \nonumber
\end{align}
This expression deals with the coordinate of $\beta$-function's zero in the special renormalization scheme:
\begin{align}
& g_* \equiv \pi^d (H(a,a,\alpha))^2 \sqrt[3]{u} = \frac{4}{3} \varepsilon - \frac{428}{243} \varepsilon^2 + \left(\frac{43088}{3^9} - \frac{160 \zeta(3)}{81} \right) \varepsilon^3 + \nonumber\\
& + \left( -\frac{5994238}{3^{13}} + \frac{11288 \zeta(3)}{3^8} + \frac{80 \zeta(4)}{27} - \frac{1280\zeta(5)}{3^5} \right) \varepsilon^4+{\cal O}(\varepsilon^5),\ d=6+2\varepsilon. \nonumber
\end{align}
The first term is invariant:
$ g_* = \frac{4}{3} \varepsilon + ...$ \ is true in any scheme. Next terms depend on scheme.

\section{Conclusions.}

Using the conformal bootstrap technique we have computed the critical index $\eta$ of $\varphi^3$ theory in 4-loop approximation analytically. Three-loop analytical result \cite{10,11} and four-loop numerical one \cite{12} were obtained earlier by using the renormalization group equation.
The evident advantage of the conformal bootstrap method is significant reduction of the number of Feynman diagrams required to obtain result. However, this technique has disadvantages --- it is applicable for the models with triple bar vertexes and conformal invariance in critical regimes only.
Though scientists managed to apply this method for calculation of the index $\eta$ of $(\varphi^2)^2$ theory within $1/n$-expansion \cite{5}, one failed to do this for construction of $\varepsilon$-expansion.

It would be interesting to modify the conformal bootstrap method for $\varphi^3$ theory so that one could apply it directly in the logarithmic dimension ($d=6$) where conformal invariance is knowingly violated. Possible physical applications of the $\varphi^3$ model are the following: first kind phase transitions \cite{13}, critical behavior near the the Yang-Lee edge \cite{14,15}. There are some interesting formal aspects of cubic models \cite{16}.

One can hope that these techniques will be useful for investigation of infrared asymptotic for Yang-Mills theory of gauge fields and describing the behavior of quarks and gluons at low energy.

\section{Acknowledgments.}

This work was supported by the SPbSU grant N 11.38.660.2013. The author is grateful for helpful conversations and correspondence with Yu. M. Pis'mak. The discussions with N. V. Antonov, L. Ts. Adzhemyan and M. V. Kompaniets were also useful.

\end{document}